\documentclass[12pt]{article}
\usepackage{epsf,amsmath,amssymb}

\newcommand{\bee}{\begin{equation}}
\newcommand{\bea}{\begin{eqnarray}}
\newcommand{\s} {\vec{s}}
\newcommand{\R}{\mathbb{R}}
\newcommand{\Z}{\mathbb{Z}}
\newcommand{\T}{\mathbb{T}}
\newcommand{\cP}{\cal P}
\newcommand{\cS}{\cal S}
\newcommand{\nonum}{\nonumber \\[1.5mm]}
\def\la{\langle}
\def\ra{\rangle}
\begin{document}                                                                
\begin{flushright}
\end{flushright}
\begin{center}
\Huge{The case against asymptotic freedom 
}\footnote{Talk presented in the Seminar at RIMS of Kyoto University: 
Applications of RG Methods in Mathematical Sciences, Sep. 10 to 12, 2003}
\end{center}
\centerline {Erhard Seiler}
\vglue 5pt
\centerline{\it Max-Planck-Institut f\"ur Physik}
\centerline{\it (Werner-Heisenberg-Institut)}
\centerline{\it F\"ohringer Ring 6, 80805 Munich, Germany}
\centerline{\it e-mail: ehs@mppmu.mpg.de}
\bigskip \nopagebreak
\begin{abstract}  \noindent
In this talk I give an overview of the work done during the last 15 
years in collaboration with the late Adrian Patrascioiu. In this work we
accumulated evidence against the commonly accepted view that theories 
with nonabelian symmetry -- either two dimensional nonlinear $\sigma$ 
models or four dimensional Yang-Mills theories --  have the property of 
asymptotic freedom (AF) usually ascribed to them.
 \end{abstract}
\vskip2mm
\section{Introduction}

Our present view of nature is based on reductionism: bulk matter consists 
of atoms, which consist of electrons and nuclei; nuclei consist of 
protons and neutrons, which in turn consist of quarks and gluons. This 
simple idea that something `consists' of something smaller becomes 
increasingly inadequate as we go down; certainly the statement that the 
proton consists of three quarks has to be taken with a big grain of salt: 
depending on the energy applied, it also seems to consists of three quarks 
and an arbitrarily large number of quark-antiquark pairs as well as 
gluons. But most of all, all these `constituents' do not exist as 
particles in the usual sense: they cannot be isolated and do not fit 
Wigner's definition of a particle as the embodiment of an irreducible 
representation of the Poincar\'e group.  
 
The catchword to describe the last property  is `confinement' and the 
theory that is supposed to embody it is Quantum Chromodynamics (QCD), 
based on Yang-Mills theory, enriched with fermion (quark) fields. Lattice 
gauge theory at strong coupling shows indeed this property of confinement 
and a huge body of lattice simulations has given us confidence that QCD 
has a very good chance to describe correctly the spectrum of hadrons. Of 
course the lattice is an artefact and one would like to get rid of it by 
taking the so-called continuum limit. The way to do this is to search for 
a critical point of the lattice theory, in which the dynamically 
generated length scale diverges in lattice units.

It is part of the conventional wisdom and stated in numerous textbooks 
that this critical point is located at vanishing (bare) coupling, and 
that the continuum limit hence enjoys a property called asymptotic 
freedom (AF), meaning that the effective interaction strength goes to 
zero with increasing energy. This prediction is based on computations 
done in perturbation theory (PT), but so far a proof does not exist.

Similar predictions have been made about two-dimensional nonlinear 
$\sigma$ models with nonabelian symmetry group, which in many ways
can be considered as toy models for QCD. But it should be stressed that 
even for this toy version AF has neither been proved nor disproved 
mathematically -- to do so remains an important challenge for mathematical 
physics.

In fact Patrascioiu and the author have questioned the textbook wisdom and 
over the years we have accumulated a lot of evidence, both analytic and 
numerical, in support of a conjecture contradicting the conventional 
picture, namely that both four-dimensional Yang-Mills theory and its 
two-dimensional toy analogues, considered as lattice statistical systems, 
have a critical point at a finite value of the lattice coupling strength, 
which then corresponds to a non-Gaussian fixed point of the  
Renormalization Group (RG). 
 
In this lecture I will focus on the two-dimensional toy version, which is 
easier to study. Concretely I will consider $O(N)$ models (classical spin 
models) which are defined as follows: we consider configurations of 
classical spins $\vec s$ on a simple square lattice $\Z^2$, i.e. maps
\bee
\Z^2\ni x \mapsto \s(x)\in \R^N,\,\,\|\s(x)\|=1\,;
\end{equation}
for any finite subset $\Lambda$  of the lattice $\Z^2$ a Hamiltonian is 
given by 
\bee
H_\Lambda=-\sum_{\la xy\ra} \s(x)\cdot \s(y)\,;
\end{equation}
On the configurations restricted to $\Lambda$,  $H_\Lambda$ induces a 
Gibbs measure
\bee
\frac{1}{Z_\Lambda}e^{-\beta H_\Lambda}d\mu_0(\{\s\})\,.
\end{equation}
As usual, one has to take the thermodynamic limit $\Lambda\nearrow\Z^2$.
The mass gap is then defined as the inverse correlation length
\bee
m(\beta)=\xi^{-1}=-\lim_{|x|\to\infty}\frac{1}{|x|}\ln \la 
\s(0)\cdot\s(x)\ra\,.
\label{mass}
\end{equation}

The textbooks state that there is a fundamental difference between the 
models with $N=2$ (abelian symmetry) and $N>2$ (nonabelian symmetry),
namely, whereas for $N=2$ we have
\bee
m(\beta)=0\,\,{\rm for}\  \beta\ge\beta_{\rm KT}\,,
\end{equation}
for $N>2$ on the contrary
\bee
m(\beta)>0 \,\,{\rm for\ all}\  \beta>0\,.
\end{equation}

The first part of the statement, which goes back to the seminal paper of 
Kosterlitz and Thouless \cite{KT}, has in fact been proven by Fr\"ohlich 
and Spencer \cite{FS} long ago. The second part, on the other hand, 
remains unproven to date and it represents an important open problem of 
mathematical physics to either prove and disprove it. Its importance for 
the understanding  of two-dimensional ferromagnets is obvious, but maybe 
even more important is its analogy to the problem of mass generation in four 
dimensional Yang-Mills theory or QCD. QCD as a valid theory of strong 
interactions requires a mass gap due to the short range nature of the 
nuclear forces, and it needs a mass scale (`string tension') describing 
the strength of the confining force.


\section {Lattice construction of Quantum Field Theories}

Let me briefly revisit the principles of constructing a massive 
(euclidean) Quantum Field Theory as a continuum limit of a lattice 
statistical model.

We assume that the thermodynamic limit of a theory on the lattice $\Z^D$ 
has been taken already and yields an infinite volume translation 
invariant Gibbs state. In the high temperature (strong coupling) regime 
$\beta<<1$ there is exponential clustering as can be shown using 
convergent cluster expansions. So the system produces its own dynamical 
scale, the correlation length $\xi$ defined in (\ref{mass}). 

We now change scale, using $\xi$ as the standard of length; so the lattice 
appears now as the rescaled version of $\Z^D$, namely $a\Z^D$ with 
$a=1/\xi$. I want to stress that therefore the {\bf lattice spacing is not 
a freely chosen parameter of the model, but a dynamically determined 
quantity}. 

In particular a continuum limit $a\to 0$ requires $\xi\to\infty$, i.e. it 
requires the existence of a critical point. By taking the correlation 
length as the standard of length, one will produce a {\it massive} 
continuum limit (mass 1 by our choice).

Returning now to the $2D$ $O(N)$ models, the procedure means that the 
continuum limit of the spin correlation functions is given by
\bee
S_{a_1,\ldots a_n}(x_1,\ldots,x_n)
=\lim_{\beta\to\beta_c}Z(\beta)^{-n/2} \la s_{a_1}(x_1\xi),\ldots, 
s_{a_n}(x_n\xi)\ra
\,.
\end{equation}
where $\beta_c$ denotes the critical value of $\beta$ (finite or infinite) 
and $Z(\beta)$ is a field strength renormalization that has to go to 0 for 
$\beta\to\beta_c$, if the continuum limit is to be a nontrivial Quantum 
Field Theory -- because in the continuum limit the correlation functions 
(Schwinger functions) has to blow up at coinciding points. A simple 
possible choice for $Z(\beta)$ is
\bee 
Z(\beta)=\frac{\chi}{\xi^2}\,,
\end{equation}
(cf. for instance \cite{PSprd01})
where $\chi$ is the susceptibility of the spin model:
\bee
\chi=\sum_{x\in\Z^2}\la \s(0)\cdot\s(x)\ra\,.
\end{equation}

While the construction of the continuum limit works whether $\beta_c$ is 
finite or infinite, a finite value for the $O(3)$ model has the crucial 
consequence that it is not aymptotically free:
\bee
\beta-c<\infty \Rightarrow {\rm No\ \  AF}\,!
\end{equation}
This follows from the bounds in \cite{PSpre98} and as has been stressed 
in \cite{PSjsp02}.

It should be noted that the construction of the continuum limit in $4D$ 
Yang-Mills theory follows the same principle; instead of the correlation 
length $\xi$ one may take $1/\sqrt{\sigma}$ where $\sigma$ is the 
string tension giving the slope of the confining potential -- it is 
believed that these two scales are equivalent, i.e. their ratio tends to a 
finite value as the continuum is approached. In any case, with this choice 
of scale (assuming it exists) one obtains a confining continuum limit. In 
this sense one may say that the `confinement problem' has been solved by 
lattice gauge theory.

In all these cases ($2D$ nonlinear $\sigma$ models and $4D$ gauge 
theories) the continuum limit has {\it no free parameter} corresponding 
to a choice of the coupling constant, there is only a scale (multiple of 
$\xi$) that may be chosen. This is the famous {\it `dimensional 
transmutation'} that is normally ascribed to properties of the 
perturbation theory for these models; we see that it is a much simpler and 
more general fact.  

\section{Arguments for Asymptotic Freedom}

All arguments are based on perturbation theory (PT). This is a formal low 
temperature expansion in powers of $1/\beta=g$. For the $O(N)$ models it 
predicts the flow of the running coupling to be governed by the 
Callan-Symanzik $\beta$ function
\bee 
\beta_{\rm CS}(g)=-\frac{N-2}{(2\pi)^2}-\frac{N-2}{(2\pi)^3}+O(g^3)\,,
\label{CS}
\end{equation}
see \cite{BZ77}).

PT is in principle just the application of Laplace's method for the 
derivation of asymptotic expansions for sharply peaked integrands. But one 
should be wary because we are dealing with infinite systems and the usual 
theorems (see for instance \cite{copson, olver}) do not apply. The idea  
is the following: the Gibbs factor $\exp(-\beta H_\Lambda)$ for $\beta>>1$ 
will be sharply peaked around the ground state configuration of 
$H_\Lambda$. The ground state of an $O(N)$ model is simply any 
configuration in which all spins ar equal:	
\bee
\uparrow\,\uparrow\,\uparrow\,\uparrow\,\ldots\, \uparrow\,\uparrow\,
\end{equation}
The expansion is produced by treating the configurations as small 
fluctuations around such a ground state. For a finite system ($L<\infty$) 
this procedure is clearly correct; but in an infinite system ($\L=\infty$)
there will {\it always} be large fluctuations according to the 
Mermin-Wagner theorem \cite{MW,DS,P}. So this means that PT should be
considered as a priori dubious.

Traditionally one has simply done the expansion in a finite volume (or 
with some other infrared regulator) and then considered the limit of the 
expansion coefficients as termwise as the the thermodynamic limit is 
taken (or the regulator is removed). In 1980 Bricmont et al \cite{bric} 
actually showed that for invariant observables in the $O(2)$ model, 
this formal procedure produces a valid asymptotic expansion for 
expectation values in the infinite volume. Nothing of that sort has been 
achieved for $N>2$.

But confidence was boosted by the finding of Elitzur \cite{eli} and later 
David \cite{david} that for $N>2$ the termwise limit exists. This is 
generally (but without justification) assumed to mean that one obtains 
indeed the correct low temperature asymptotic expansion by PT.

But the problem cannot be settled by looking at the termwise limits. The 
structure of the problem is as follows: let $A$ be some observable; 
then for $L<\infty$ we have
\bee
\la A\ra_L=\sum_{n=0}^M c_n(L)\beta^{-n}+ R_M(\beta,L)
\end{equation}
with
\bee
\vert R_M(\beta,L)\vert\le K_M(L)\beta^{-M-1}\,.
\end{equation}
Elitzur and David showed that $\lim_{L\to\infty}c_n(L)$ exists, but to 
justify PT one would have to prove uniform bounds on $K_M(L)$.

\section{Reasons for doubt}
I list a few facts that should be reason not to trust PT in the $O(N)$ 
models for $N\ge 3$,

\begin{itemize}
\item
I mentioned already that the Mermin-Wagner theorem requires large 
fluctuations at all temperatures and so destroys the logical basis of PT.

The intuitive reason for this fact lies in the existence of localized 
defects of arbitrarily low energy, which disorder the system at all 
temperatures. We dubbed these defects `superinstantons', because they 
will `beat' the instantons (which have fixed energy) at low temperature. 
They can be simply described as follows: fix a spin at the origin to a 
certain direction, say $\vec e_N$, the unit vector in the $N$th 
direction; also fix all spins at approximate distance $R$ from the origin 
to another direction $\vec e'$. The configuration of minimal energy 
with these boundary conditions (bc) is called superinstanton of size $R$ 
(and rotation angle $\arccos(\vec e_N\cdot\vec e')$). It turns out that 
the energy 
of such a superinstanton is 
\bea 
E(R)=O(R^{-1}) \ {\rm in} \ D=1 \nonum
E(R)=O(1/\ln R)  \ {\rm in} \ D=2 \nonum
E(R)=O(R^0) \ {\rm in} \ D>2 
\end{eqnarray}

(see for instance \cite{SY}). A notable property of these excitations is 
that they do not exploit the fact that the spin has $N$ components; they 
are $O(2)$ like for all $N$ and all $D$. They are local minima of the 
energy (with the prescribed bc); in the continuum the $O(2)$ 
superinstanton can be seen to be the conjugate harmonic function to the 
vortex that plays such a fundamental role in the Kosterlitz-Thouless 
(KT) theory.
 
\item
Ambiguity of PT \cite{PSprl95}:

The thermodynamic limit of $2D$ $O(N)$ models ought to be independent of 
bc. This is plausible because of the Mermin-Wagner theorem, but it is a 
definite fact in the massive phase (which for $N>2$ according to the 
conventional wisdom is all there is) because of exponential clustering.

Nevertheless Patrascioiu and the author found some bc that lead to PT 
coefficients that differ from the standard ones (obtained with periodic 
bc) in the {\it termwise} thermodynamic limit. These bc were inspired by 
the superinstantons and called superinstanton bc (sibc). They are defined 
essentially by enforcing a superinstanton of rotation angle zero. If we 
consider $O(N)$ invariant observables, sibc can be defined by fixing the 
spin at the origin and all the spins at the boundary of a square to a 
certain fixed direction. 

It is inportant to realize that these are indeed legitimate bc for the 
model: in the thermodynamic limit the effect of the boundary spins 
disappears, and fixing one spin in the center is inconsequential, since we 
are considering invariant observables.   

For $N>2$ (but not for $N=2$!) PT at one-loop level (termise 
thermodynamic limit), already produces an answer different from the 
standard one if one uses sibc~. This means that at least one of the two 
results, possibly both, are incorrect, since the true (nonperturbative) 
thermodynamic limit is the same and an asymptotic expansion is unique, if 
it exists.

\item
Richard's truncated sphere model \cite{richard}:

If one modifies the $O(3)$ model by restricting the $z$ component of the 
spin to satisfy the constraint $|\s_z|<1-\epsilon$, the system becomes 
massless at sufficiently low temperatures. This was shown by J.L.Richard 
by the use of Ginibre's correlation inequalities for the $x,y$ components 
and comparing the model to an $O(2)$ model. 

But PT would tell a different (and untrue) story: PT, expanding 
around a ground state oriented towards the equator does not {\it see} the 
constraint at all. It yields the Callan-Symanzik $\beta$ function of the 
ordinary $O(3)$ model (\ref{CS}) and therefore, by the usual reasoning, 
says that the model is aymptotically free and massive for all $\beta$! 

\end{itemize}

\section{The case against AF: discrete and continuous models}

It is interesting to make a `lattice approximation of the target space' by 
replacing for instance the 2-sphere $S_2$ of the $O(3)$ model with (the 
set of vertices of) a platonic solid such as the icosahedron or 
dodecahedron. In both of these cases the symmetry is also reduced from the 
continous $O(3)$ to the discrete icosahedral subgoup $Y$. 

The study of the dodecahedron model was begun already long ago, both 
analytically and numerically in \cite{PRSplb90, PRSplb91} and continued in 
\cite{PSplb98}; a detailed numerical study of the icosahedron model was 
carried out in \cite{PSprd01}. The picture that emerged is that the 
dodecahedron has a critical interval between its massive high and low 
temperature phases (similar to the abelian $Z_n$ models for large enough 
$n$), whereas the icosahedron model has an isolated critical point 
$\beta_c$. In both cases our main interest is in the (massive) continuum 
limit arising by approaching criticality  from the high temperature phase.

Numerically we found evidence for symmetry enhancement from $Y$ to $O(3)$ 
in both cases, suggesting that the massive continuum limits of the 
discrete models lie in the same universality class as the $O(3)$ model.

Concretely, we considered the following objects:

\begin{itemize}
\item
Renormalized spin-spin correlation in momentum space
\bee
G_r\left(\frac{n}{\xi}\right)\equiv \frac{\xi^2}{\chi} \la \vec
s(0)\cdot\vec s(n)\ra\,,
\end{equation}
where the data presented in \cite{PSplb98} suggest
\bee 
\boxed{\lim_{\xi\to\infty} \hat G_r(p)\vert_{\rm dodeca}=
\lim_{\xi\to\infty} \hat G_r(p)\vert_{O(3)}}\,.
\end{equation}

\item
Renormalized spin-spin correlation in $x$ space
\bee
G_r\left(\frac{n}{\xi}\right)\equiv \frac{\xi^2}{\chi} \la \vec
s(0)\cdot\vec s(n)\ra\,,
\end{equation}
where the data presented in \cite{PSplb02} suggest
\bee
\boxed{\lim_{\xi\to\infty}\hat G_r(x)\vert_{\rm ico}= \lim_{\xi\to\infty}
G_r(x)\vert_{O(3)}}\,.
\end{equation}

\item 
Renormalized 4-point coupling constant
\bee
g_R=\left(\frac{g_4}{g_2^2}-\frac{5}{3}\right) \frac{\xi^2}{\chi}\,,
\end{equation}
where the data in \cite{PSplb02} give evidence for 
\bee
\boxed{\lim_{\xi\to\infty}
g_R\vert_{\rm ico}=g_R\vert_{O(3)}}
\end{equation}
\end{itemize}

This suggests the overall conclusion that the dodecahedron, the 
icosahedron and the $O(3)$ model all have the same massive continuum 
limit. This is relevant for our main issue, because the discrete models 
cannot be expected to be asymptotically free. In fact the running coupling
proposed by L\"uscher, Weisz and Wolff \cite{LWW}
\bee
\overline{g}_{LWW}(z)\equiv \frac {L}{\xi(L)} \qquad
\left(z=\frac{L}{\xi(\infty)}\right)
\end{equation} 
in the icosahedron model has found to have an ultraviolet fixed point 
\cite{PSplb02}  
\bee
\overline{g}^\ast\approx .595 \ne 0\,.
\end{equation}

So assuming the universality suggested by our does indeed hold, this 
means that the $O(3)$ model cannot have AF.

\section{The case against asymptotic freedom: percolation}

The strongest argument against AF developed by Patrascioiu and the author 
is based on the analysis of certain percolation properties. The idea was 
presented first already in 1991 and developed in a number of papers 
\cite{PSjsp92, PSprl92, PSlat92, PScrm93}. Since the argument does not 
constitute a rigorous proof, a detailed numerical study to bolster it was 
carried out much later \cite{PSjsp02}.

The argument might seem tricky, but it involves some solid analytic 
results, on which the heuristic/numerical arguments builds.

\begin{itemize}

\item
The first step are two modifications of the model that should not change 
its universality class:
\begin{itemize}
\item
We replace the square lattice $\Z^2$ by a triangular lattice $\T$; this is 
achieved simply by adding extra bonds along one of the diagonals of each 
elementary square. All the bonds carry the same Gibbs factor, so the 
model is really living on an isotropic triangular lattice. There is no 
question that this modification does not affect the continuum limit, i.e. 
does not change the universality class. 
\item
Next we introduce a constraint in the Gibbs measure that limits the angle 
between neighboring spins to some maximal value:
\bea
e^{-\beta H}\ \  &&\rightarrow \ \  e^{-\beta H}
\prod_{\la xy\ra} \theta(\vec s(x)\cdot \vec s(y)-c)\nonumber\\
\mbox{(`standard action')}\ \  &&\rightarrow
\mbox{\ \  (`cut action')}
\end{eqnarray}

For large values of $\beta$ this change is completely innocuous, because 
the standard action Gibbs factor will already make large angles extremely 
unlikely. Our arguments will be completely independent of the value of 
$\beta$, so one might assume that it is already fixed at some large value. 
It turns out, however, that one can change the point of view by putting 
$\beta=0$ and varying the cut parameter $c$ instead, since the presence of 
the constraint also has a ferromagnetic ordering effect of the system.
One finds indeed that the correlation length increases with increasing $c$
and a critical value of $c$ somewhere near 0.76 seems to exist.

We also performed a direct numerical test of universality between this 
`cut' model with $\beta=0$ and the standard action model by measuring the 
so-called `step scaling function' that gives the change of the LWW 
coupling under doubling the scale as a function of the LWW coupling 
itself. The data show quite good agreement between the two models
(see \cite{PSjsp02}).
\end{itemize}
\item
An Ising model is imbedded in the $O(3)$ model by setting
\bee
\sigma_x\equiv {\rm sgn}\, s_z(x)
\end{equation}
as is done for the well-known cluster algorithms. Note that the definition 
of the Ising spins is well-defined except on a set of measure zero.  

\item
A correlated bond percolation model is set up following the original work 
of Fortuin and Kasteleyn \cite{FK} and adapted to $O(N)$ models by Wolff 
\cite{wolff}:

Given a spin configuration, bonds (nearest neighbor pairs) are activated 
with a conditional probability $p(\la xy \ra| \{\s(x)\, x\in\T\}$ that is 
determined by the change in the Gibbs factor when the $z$ component of 
one of the spins is reflected: 
\bee
p(\la xy \ra| \{\s(x)\, 
x\in\T\}=\theta(\sigma_x\sigma_y)\left[1-e^{\left(\beta 
\s(x)\cdot\left(R_z\s(y)-\s(y)\right)\right)}\right]
\end{equation}
where
\bee
\vec s\mapsto R_z\vec s=\s-2\vec e_z (\vec e_z\cdot \s)
\end{equation}
After activating the $\la xy\ra$ bonds independently with this probablity 
one forms connected bond clusters; then one averages over all spin 
configurations with the Gibbs measure.

A theorem that is essentially due to Fortuin and Kasteleyn \cite{FK} now 
relates the expected size $\la C_{FK}\ra$ (i.e. number of vertices) of
the bond cluster attached to the origin to the susceptibility of the 
imbedded Ising spins:
\bee
\chi_{\rm Ising}=\la C_{FK}\ra\,.
\end{equation}
 
This theorem has an important corollary:
\bee
\boxed{{\rm If}\ \  \la C_{FK}\ra=\infty,{\rm  then}\ \  \xi=\infty}.
\end{equation}

\item
The next step is to divide the 2-sphere into 3 regions:
\begin{itemize}
\item
Equatorial strip $\cS_\epsilon\equiv$ $\{\vec s\in S^2 |\, |s_z|\le
\frac{\epsilon}{2}\}$
\item
North polar cap $\cP_\epsilon^+ \equiv$ $\{\vec s\in S^2 | s_z>
\frac{\epsilon}{2}\}$
\item
South polar cap $\cP_\epsilon^- \equiv$ $\{\vec s\in S^2 | s_z<
-\frac{\epsilon}{2}\}$

\end{itemize}

The idea is now to study clusters of 
$\cS_\epsilon,\  \cP^+_\epsilon\cup\cP^-_\epsilon\equiv\cP_\epsilon$.

The activation probability for the bond $\la xy\ra$ is always 1 if 
$c>1-\frac{\epsilon^2}{2}$ and $\s(x),s(y)\in \cP^+_\epsilon$ (or 
$\s(x),s(y)\in \cP^-_\epsilon$), because in this case flipping one of the 
spins would violate the constraint. So we have
\bee 
\la C_{FK}\ra\ge \la\cP^+_\epsilon\ra
\end{equation}

\item
The main result of \cite{PSjsp92,PSlat92} is the following:

{\bf If for some $\epsilon$ and some $c>1-\epsilon^2/2$
\ \ $\cS_\epsilon$ does NOT percolate, then $\xi=\infty$.}

The idea of the proof in intuitive terms is as follows:
assume that $cS_\epsilon$ does not percolate, then there are two 
possibilities:
\begin{itemize}
\item 
$\cP_\epsilon$ percolates $\Rightarrow$ $\cP_\epsilon^+,\
\cP_\epsilon^-$ percolate: but this is impossible because in $2D$ there 
cannot be two disjoint percolating clusters (actually this is a 
`principle', proven only for Bernoulli percolation and the Ising model)
\item 
$\cP_\epsilon$ does not percolate, but prevents percolation of
$\cS_\epsilon$

$\Rightarrow$ $\cP_\epsilon^+,\  \cP_\epsilon^-$ form `rings' of 
arbitrarily large size; neither $\cP_\epsilon^+$ nor its complement 
percolate. Now by a lemma of Russo \cite{russo}, if on a self-matching 
lattice (such as $\T$) neither a set nor its complement percolate, the
expected cluster size of both of them diverges: 
$\la \cP_\epsilon^+\ra=\infty\ \Rightarrow
\la C_{FK}\ra=\infty \ \Rightarrow \xi=\infty$, which completes the 
argument.

\end{itemize}

\item
So the question remains whether it is possible for an equatorial strip 
$\cS_\epsilon$ to percolate for arbitrarily small $\epsilon$ and 
$c>1-\frac{\epsilon^2}{2}$. This is a priori hard to imagine. In any case, 
in \cite{PSlat92, PScrm93} we gave an argument that leads to masslessness 
even if we assume this implausible situation to occur:

Assume $\cS_\epsilon$ percolates. Then the same is of course true for any
$\cS_{\epsilon'}$ with  $\epsilon'>\epsilon$. Taking $\epsilon'$ close to 
2 one obtains an `ocean' with only rare and small islands corresponding 
to $\cP_\epsilon'$. Focussing on the two components $s_x, s_y$, and
taking $\beta$ very large, we have a low temperature $O(2)$ model with 
fluctuating coupling on that ocean. One can arrange for the islands to 
cover an arbitrarily small fraction of the lattice and at the same time 
make the effective temperature for the $O(2)$ model arbitrarily small. So 
one expects that one ends up in a massless Kosterlitz-Thouless (KT) phase. 
The only problem is that the Fr\"ohlich-Spencer proof has not been 
adapted for this situation. 

Of course I should stress that I don't think this situation ever arises; I 
rather expect that at low enough temperature (large enough $c$) the 
clusters of an arbitrarily small polar cap form rings of any size and have 
divergent cluster size. This conforms to the superinstanton picture and
parallels the situation rigorously established for the `cut' $O(2)$ model 
(see \cite{PSprl92,aiz}). 

\item
Finally I want to briefly describe the extensive numerical study of the 
percolation properties in \cite{PSjsp02}:

For simplicity we studied only the model with $\beta=0$, but varying cut 
$c$. A nonzero $\beta$ would only order the system more; if it is massless 
at $\beta=0$ it would a fortiori be so for $\beta>0$. We scanned the 
$(c,\epsilon)$ plane and measured the ratio  
$r(L)=\la\cP_\epsilon\ra/\la\cS_\epsilon\ra$ as a function of the lattice 
size $L$. For fixed $0.78<\epsilon<1$ and varying $c$ betwqeen -0.1 and 
0.9 three regions could be distinguished:
\begin{itemize}
\item
For small $c$, $r(L)$ increases rapidly with $L$, indicating percolation 
of $\cP_\epsilon$ -- it should be noted that in this regime the clusters
of $\cP^+_\epsilon$ and $\cP^-_\epsilon$ are allowed to touch.
\item
For intermediate $c$, $r(L)$ decreases rapidly with $L$, presumably 
converging to 0 and indicating percolation of $\cS_\epsilon$. The boundary 
between these two regions is a rather sharply defined value of $c$, where 
$r(L)$ is practically independent of $L$
\item
for even larger $c$, this $L$ dependence of $r(L)$ seems to disappear,  or 
even change direction again, but at most showing a very mild increase with 
$L$ (maybe powerlike). We take this as an indication that now neither of 
the sets $\cP\epsilon$, $\cS_\epsilon$ percolates (which means by Russo's 
lemma that they both have divergent mean cluster size).

\end{itemize}
It turns out furthermore that for $c<0.76$ or so the intermediate region 
of percolating $\cS_\epsilon$ seems to disappear altogether. To assess the 
credibility of these conclusions, one should note that we went to quite 
large lattices (up to $L=1280$), but numerical results are of course never 
a substitute for a proof. 

We can combine the picture suggested by the numerical results into a 
(semiquantitative) `percolation phase diagram' (Fig. \ref{phase}) taken 
from \cite{PSjsp02}.
 
\begin{figure}[htb]
\centerline{\epsfxsize=9.0cm\epsfbox{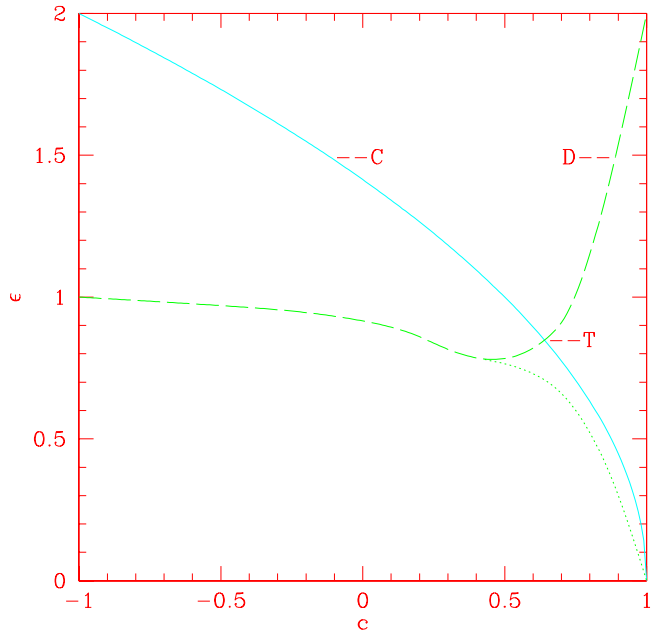}}
\caption{Percolation phase diagram of the $O(3)$ model on the $\T$ 
lattice. C is the line $c=1-\epsilon^2/2$; above the dashed line D 
$\cS_\epsilon$ percolates, above the dotted line $\la 
\cS_\epsilon\ra<\infty$.}
\label{phase}
\end{figure}

To corroborate our conclusion that for $c<0.76$ neither $\cP\epsilon$ 
nor $\cS_\epsilon$ percolates, we carried out a different analysis of the 
equatorial clusters with width $\epsilon=0.75$; this time we compared them 
to the clusters of $\cP^+_{\epsilon'}$, chosen such that the polar cap has 
the same area as the equatorial strip $\cS_\epsilon$; also both 
sets cover the same fraction of the lattice. The ratio $r'(L)=\la 
\cP^+_{\epsilon'}\ra/\la \cS_\epsilon\ra$ is always larger than 1, 
presumably because the polar cap has less boundary than the equatorial 
strip of the same area (in agreement with a conjecture stated in 
\cite{pat}). For small $c$ (roughly up to $c=0.4$) $r'(L)$ shows a rapid 
increase with $L$, indicating that $\cP_{\epsilon'}$ forms rings (of all 
sizes) whereas the clusters of $\cS_\epsilon$ have finite mean size.
At $c\approx 0.4$ the behavior of $r'(L)$ changes: it is still growing 
with $L$, but now only powerlike. This can only indicate that both kinds 
of clusters now form rings of arbitrary size; it rules out for all 
practical purposes that $S_\epsilon$ percolates, which would require a 
decrease to 0 as $L\to\infty$. 

The numerical results of this study are shown in Fig.\ref{rings}, taken 
from \cite{PSjsp02}.

\begin{figure}[htb]
\centerline{\epsfxsize=11.0cm\epsfbox{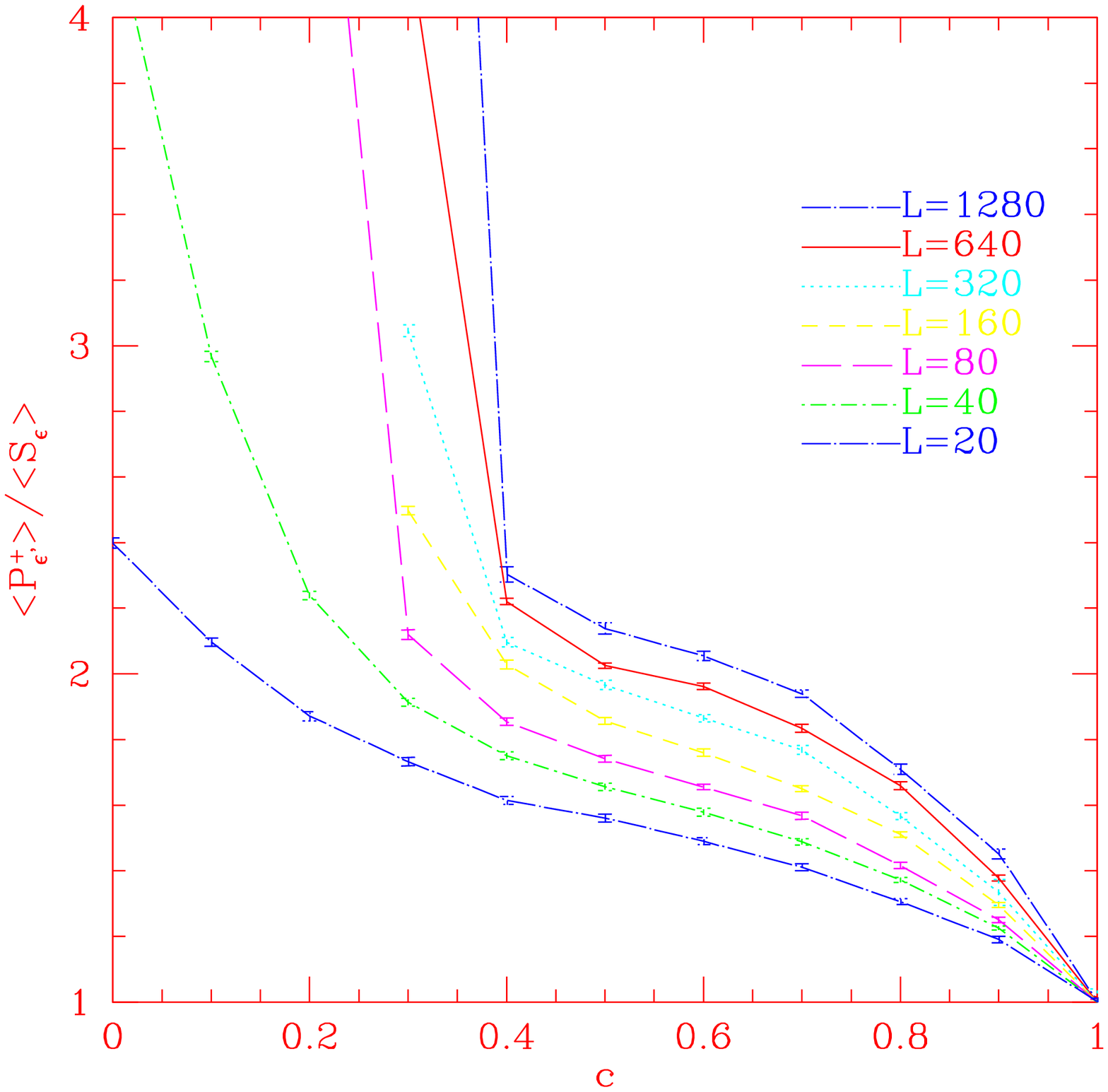}}
\caption{The ratio of the mean cluster size of
a polar cap of height 0.75 to that of an equatorial
strip of the same height}
\label{rings}
\end{figure}

To sum up: our numerical study gives strong evidence that for sufficently 
small $\epsilon$ (about $<0.76$) $\cS_\epsilon$ does not percolate for 
any value of $c$. This implies according to our results that for 
$c>1-0.76^2/2=0.745$ the correlation length $\xi=\infty$. The real 
challenge is of course to give a mathematical proof of this.
\end{itemize}

\section{Where do we stand?}

Clearly, the problem of AF is still open
\begin{itemize} 

\item
Textbook wisdom is insufficient to settle it positively
\item
Counterarguments are not rigorous
\item
Any progress towards proving or disproving AF is very desirable. For this 
reason programs such the one of K.R.Ito \cite{ito}, attempting to prove AF 
by a controlled Renormalization Group approach, are very welcome.
\end{itemize}
A final remark: Solving the problem of AF for the $2D$ toy models which 
were mostly discussed here would be a big step towards understanding
it also in $4D$ Yang-Mills theory and thereby also towards a solution of 
one of the Million Dollar Problems posed by the Clay Mathematics 
Institute.


\end{document}